\documentclass{iau} 
\usepackage{graphicx}

\title[Stellar orbits and origin of the stellar halo] %% give here short title %%
{Stellar orbital properties as diagnostics of the origin of the stellar halo }

\author[Monica Valluri et al.]   %% give here short author list %%
{Monica Valluri$^1$,  Sarah R. Loebman$^{1,2}$, Jeremy Bailin$^3$, Adam Clarke$^4$, Victor P. Debattista$^4$, Greg Stinson$^5$}

\affiliation{$^1$University of Michigan, USA email: {\tt mvalluri@umich.edu},
$^2$Michigan Society of Fellows, $^3$University of Alabama, USA, $^4$University of Central Lancashire, UK, $^5$Max Planck Institute for Astronomie, Germany}

\pubyear{2015}
\volume{317}  %% insert here IAU Symposium No.
\jname{The General Assembly of Stellar Halos: Structure, Origin and Evolution}
\editors{M. Arnaboldi, A. Bragaglia, M. Rejkuba \& D. Romano eds.}

\begin{document}

\maketitle

\begin{abstract}
We examine metallicities, ages  and orbital properties of halo stars in a Milky-Way like disk galaxy formed in the  cosmological hydrodynamical MaGICC simulations. Halo stars were either accreted from satellites or they formed \textit{in situ} in the disk or bulge of the galaxy and were then kicked up into the halo  (``in situ/ kicked-up'' stars). Regardless of where they formed \textit{both types show surprisingly similar orbital properties}: the majority of both types are on short-axis tubes with the same sense of rotation as the disk -- implying that a large fraction of satellites are accreted onto the halo with the same sense of angular momentum as the disk. 
\end{abstract}

\firstsection % if your document starts with a section,
\section{Introduction}
The orbital properties of halo stars from the MaGICC simulated galaxy g15784 -- a realistic Milky Way sized disk galaxy at $z=0$ \cite[(Stinson et al. 2012)]{stinson_etal_12} -- are used to assess whether it is possible to determine the birth site of halo stars based on individual orbital properties. The dark matter halo of this galaxy is mildly triaxial at all radii. The stars were picked to lie within 100 kpc of the center of the galaxy but not within the disk (i.e at  $|z| > 3$~kpc and $R>25$kpc). The formation sites of halo stars were determined by examining 100 snapshots of the simulation.  ``Accreted halo stars''  form in satellites outside the virial radius and are subsequently added to the halo via tidal stripping. Stars that form ``in-situ'' in the main disk or bulge and are subsequently kicked into the halo are referred to as ``in situ/ kicked-up'' stars. %(A population of halo stars which which is formed in the satellites after they have entered the Virial radius (referred to in the literature variously as ``endodebris'', ``ex-situ stars'' and ``commuter stars'') are ignored in this work since their properties are likely to be much more sensitive to feedback prescriptions and require more detailed analysis. )
The phase space coordinates of individual stars at $z=0$ were used to numerically integrate individual orbits  in the frozen potential corresponding to the full galaxy potential. Frequency analysis of the orbits was used to classify orbits \cite[(Valluri et al. 2010, 2012)]{valluri_etal_10, valluri_etal_12} into the major orbit families found in triaxial halos: box orbits, short axis tubes (SAT), long axis tubes (LAT) and chaotic orbits. 

\begin{figure}
%\begin{minipage}{\textwidth}
%\begin{minipage}{0.75\textwidth}
%\centering
\includegraphics[trim=0.pt 0.pt 0.pt 0pt,width=0.32\textwidth]{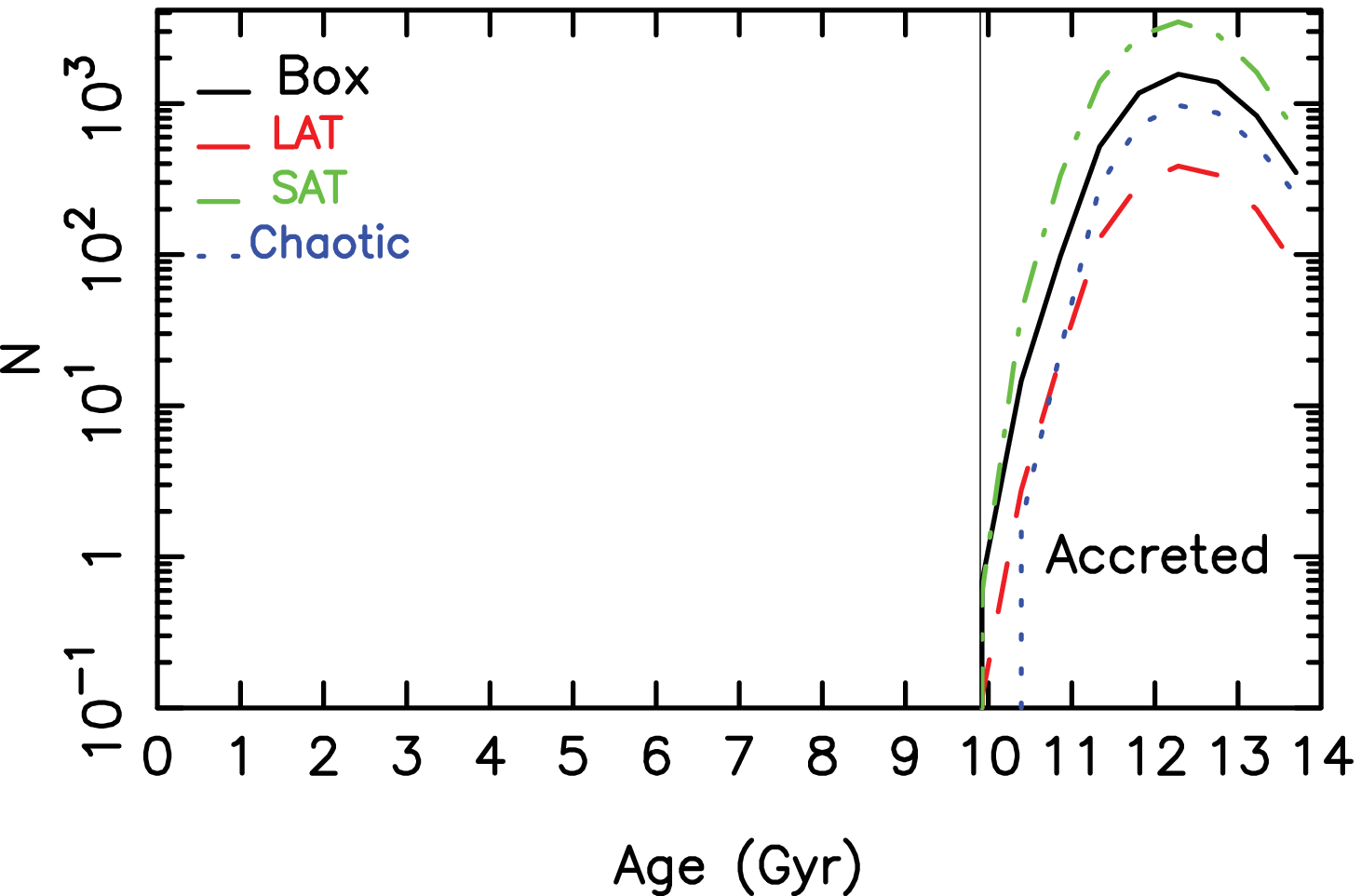}
\includegraphics[trim=0.pt 0.pt 0.pt 0pt,width=0.32\textwidth]{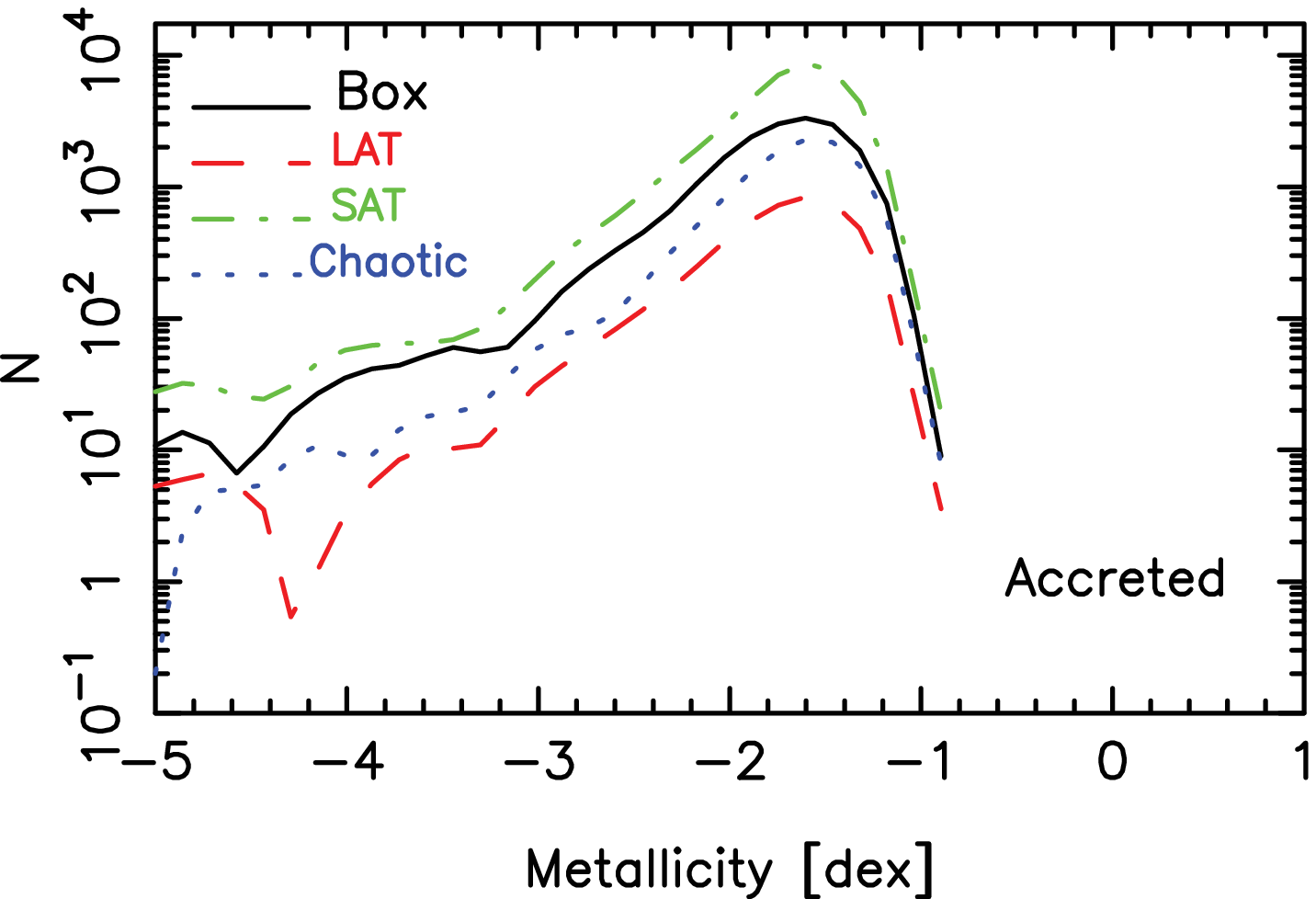}
\includegraphics[trim=0.pt 0.pt 0.pt 0pt, clip, width=0.32\textwidth]{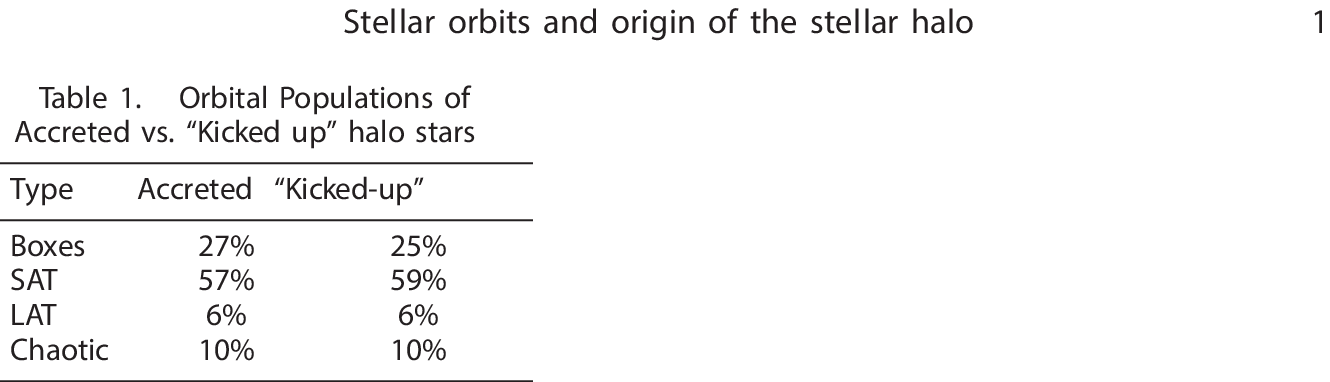}
\includegraphics[trim=0.pt 0.pt 0.pt 0pt,width=0.32\textwidth]{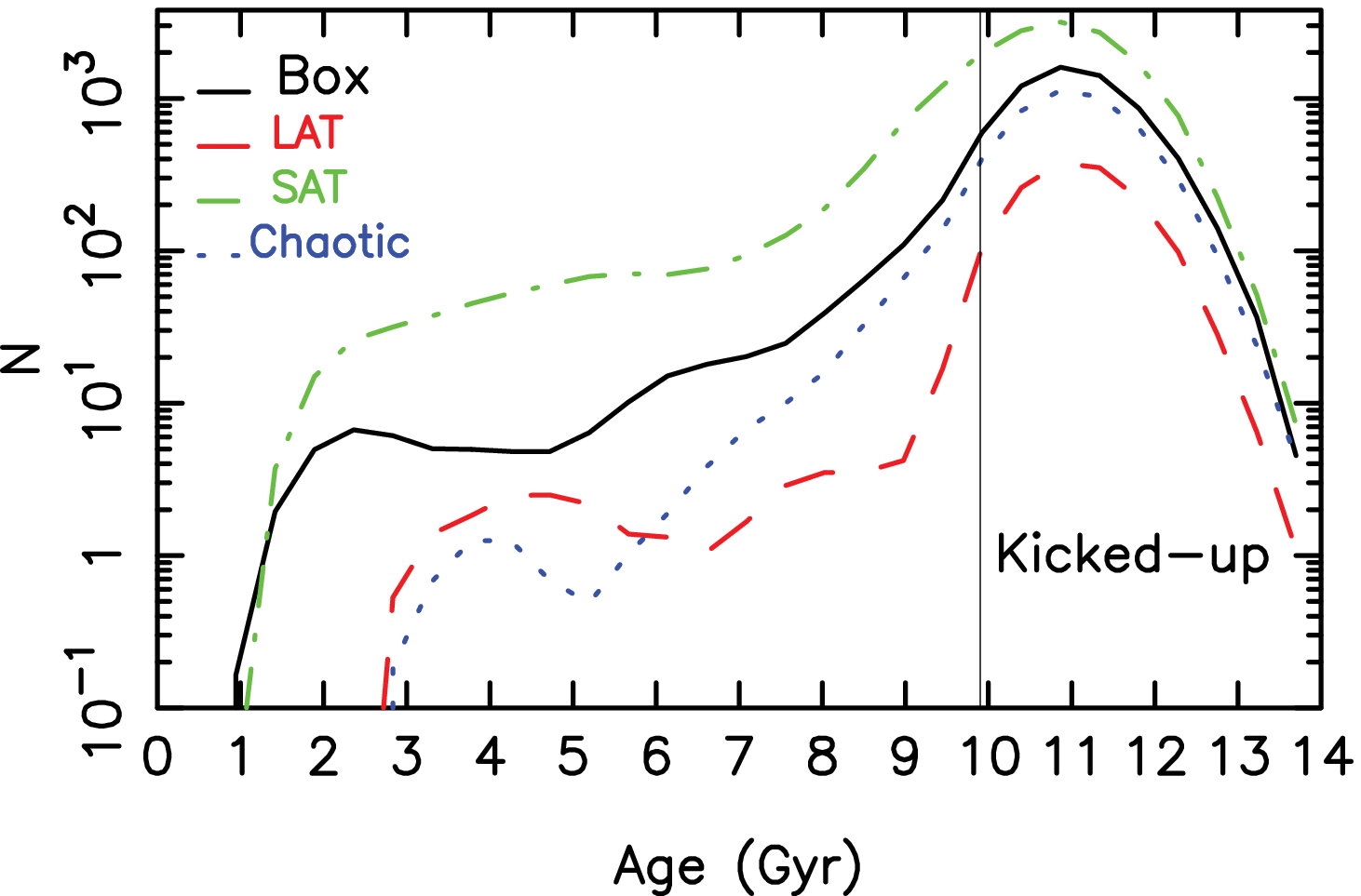}
\includegraphics[trim=0.pt 0.pt 0.pt 0pt,width=0.32\textwidth]{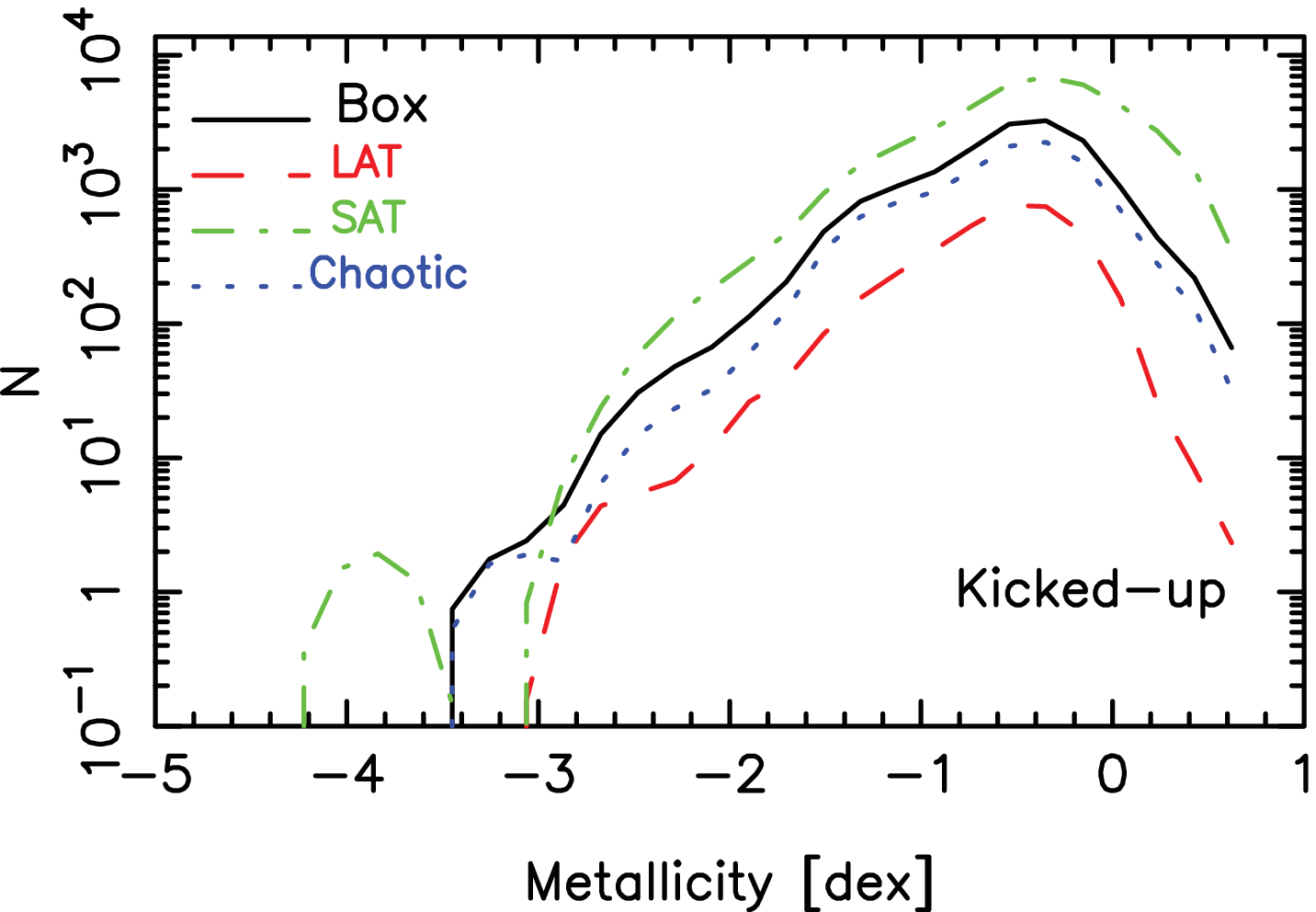}
 %\vspace*{-0.3cm}
 \caption{\footnotesize Kernel density histograms showing number of orbits of each family as a function of stellar age (left) and metallicity (right) for accreted stars (top row) and ``in situ/ kicked up'' stars (bottom row). Orbits of all families
 are similarly distributed  in each panel. The thin vertical lines (at 9.9 Gyr) marks the youngest accreted stars.
 }
\label{fig1}
%
%
%\centering
\end{figure}

%\begin{table}
%\begin{minipage}{0.40\textwidth}
%\caption{Orbital Populations of Accreted vs. ``Kicked up'' halo stars}
%\begin{tabular}{lcc}\hline 
%{\bf Type} & {\bf Accreted} & {\bf ``Kicked-up''}   \\  \hline
%{\bf Boxes}     & 27\%         &25\%\\ 
%{\bf SAT}        &  57\%        & 59\%\\
%{\bf LAT}          & 6\%           &  6\%\\
%{\bf Chaotic}    & 10\%         & 10\%\\
%\hline
%\end{tabular}
%\end{minipage}
%\end{table}
%\end{minipage}
%\end{minipage}

\vspace{-0.2cm}
\section{Orbits  of accreted and ``kicked-up'' halo stars}

Orbits of 14,000 accreted stars 14,000 ``kicked-up'' stars were classified. Table~1 shows the percentage of accreted and ``kicked-up'' stars belonging to each orbit family.   The  majority of orbits are on SATs with the same sense of rotation as the disk, regardless of where they formed. This is probably because satellies are often accreted together along the same large-scale filaments that contribute to the hierarchical growth of the galaxy \cite[(Helmi et al. 2011)]{helmi_etal_11}. More surprising is the fact that the fractions of box orbits and chaotic orbits are independent of formation site: however these orbit families are more centrally concentrated in the halo and are likely to have experienced more chaotic scattering \cite[(Valluri et al. 2013)]{valluri_etal_13}. We find that the distribution of orbital chaoticity in this high resolution simulation is identical to that in controlled simulations \cite[(Valluri et al. 2010)]{valluri_etal_10}. Figure~\ref{fig1} shows that ``accreted'' and ``in situ'' stars have slightly different distributions with age and metalliticy, but all orbit families in a given panel show similar distrbutions in these quantities. Accreted stars are older on average than ``in situ'' stars but the peak  of the ``in situ'' stars  is also old. Likewise there is significant overlap in their metallicity distributions. \cite[Snaith et al. (2015)]{snaith_15} show that while the accreted halo stars have higher $\alpha$-abundances on average than ``in situ'' halo stars, there is overlap in this space too.
\vspace{-0.2cm}

\section{Conclusions}
Orbital properties of halo stars are independent of whether they were accreted or whether they were formed in the disk/bulge and then ``kicked-up'' into the halo. Both types of stars are mainly on short-axis tubes ($\sim 60$\%), probably because the  accretion of satellites occurs on a few large scale filaments that also feed the disk. Chaotic scattering by the central bulge put stars in the inner halo on box or chaotic orbits (Valluri et al. 2010). In this MaGICC galaxy the overlap in the ages, metallicities and orbital properties of halo stars makes it difficult to uniquely identify exactly where they formed. This analysis is being repeated for other  disks to assess the dependence of this result on accretion history.

\acknowledgements MV is supported by NASA-ATP grant NNX15AK79G. SRL is supported by the Michigan Society of Fellows, VPD is supported by STFC Consolidated grant \# ST/M000877/1.  JB acknowledges support from HST-AR-12837, provided by NASA through a grant from the Space Telescope Science Institute.
{}

\end{document}